\documentclass[prl,a4paper,twocolumn,showpacs]{revtex4}
\usepackage{amssymb}
\usepackage{graphicx}

\begin{document}

\title{Gravity waves on the surface of topological superfluid $^3$He-B}

\author{V.B.~Eltsov}
\author{P.J. Heikkinen}
\author{V.V. Zavjalov}

\affiliation{O.V. Lounasmaa Laboratory, Aalto University, P.O. Box 15100,
  FI-00076 AALTO, Finland}

\begin{abstract}
  We have observed waves on the free surface of $^3$He-B sample at
  temperatures below 0.2\,mK. The waves are excited by vibrations of the
  cryostat and detected by coupling the surface to the Bose-Einstein
  condensate of magnon quasiparticles in the superfluid. The
  two lowest gravity-wave modes in our cylindrical container are
  identified. Damping of the waves increases with temperature linearly with
  the density of thermal quasiparticles, as expected. Additionally finite
  damping of the waves in the zero-temperature limit and enhancement of
  magnetic relaxation of magnon condensates by the surface waves are
  observed. We discuss whether the latter effects may be related to
  Majorana fermions bound to the surface of the topological superfluid.
\end{abstract}

\date{\today}

\maketitle

Waves on the surface of a fluid in a gravitational field \cite{LL} present
a universal phenomenon in a wide range of systems, from a
glass of drink to hot astrophysical
objects \cite{surfnstar} and cold superfluids \cite{surf4he}.  Properties of
the waves provide an important information about the fluid itself which in
turn can result in useful practical applications.  For example using the
well-known phenomenon that oil film stills the water waves \cite{oilfilm}
one can find oil pollutions in the ocean by observing its calm regions from
a satellite.

Recently the surface properties of the fermionic $^3$He in its superfluid B
phase have 
attracted a lot of attention owing to non-trivial topology
of this superfluid: It is expected that fermionic bound states with
Majorana character emerge at the surface \cite{he3majorana,topsuper,elrelax,okudarev}. While in
solid-state systems complex engineering efforts are required to obtain
Majorana fermions \cite{majoranasss}, the free surface of $^3$He-B should be naturally
covered by a thin layer of such states. Could this 'film' in a sense
'still' the surface waves of $^3$He-B and is it possible to observe this
damping in the experiment? While this question awaits a proper theoretical
consideration we report here the first, to our knowledge, observation of
gravity waves on the surface of $^3$He-B.

For such observation it is not enough to cool $^3$He below its critical
temperature $T_{\rm c} \approx 10^{-3}\,$K. At temperatures close to
$T_{\rm c}$ viscosity of the normal component of $^3$He is high, oil-like,
and the surface waves are overdamped \cite{LL}. Only the third sound waves
in a thin film,
where the normal component is clamped, have been previously observed in
$^3$He-B at $(0.3 \div 0.8)\,T_{\rm c}$ \cite{3rdsound}. We
have performed measurements at temperatures below $0.2\,T_{\rm c}$ where
the normal component becomes a rarefied gas of ballistic quasiparticles and
its contribution to the damping of the surface waves rapidly decreases.

\begin{figure}[t]
\centerline{\includegraphics[width=\linewidth]{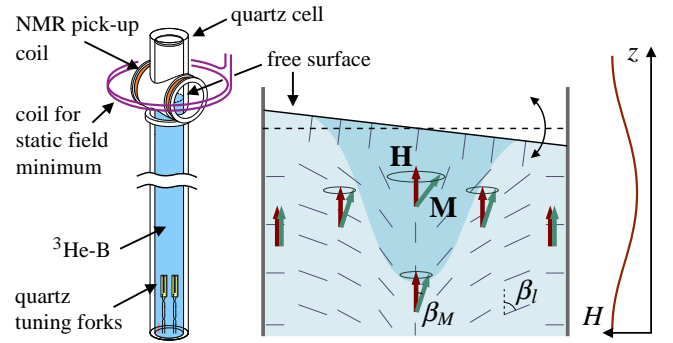}}
\caption{Experimental setup and principle of measurements. \textit{(Left)}
  Sample tube with the NMR pick-up coil around the free surface and
  tuning-fork thermometers at the bottom, where the cylinder opens to the
  heat-exchanger volume of the nuclear cooling stage. \textit{(Right)}
  Bose-Einstein condensates of magnon quasiparticles can be created in a
  trap, formed in the radial direction by texture of the the orbital
  anisotropy axis (sketched with the short segmented lines) and in the
  axial direction by the applied minimum of the static magnetic field $H$.
  In the condensate magnetization $\mathbf{M}$ precesses around the
  magnetic field with coherent phase and at the common frequency, which is
  measured in the experiment. The frequency is determined by the trapping
  potential and is modulated by the oscillations of the surface.}
\label{setupfig}
\end{figure}

\textit{Experiment.}
The $^3$He-B sample is contained in a vertical
quartz cylinder with internal diameter of $2R = 6\,$mm and length of
150\,mm, Fig.~\ref{setupfig}. The free surface of
the superfluid is placed about 8\,mm below the upper wall of the cylinder
within the pick-up coil of the nuclear magnetic resonance (NMR) spectrometer. 
To detect surface oscillations we use their influence on the frequency of
the coherently-precessing NMR mode known as the trapped Bose-Einstein
condensate of magnon quasiparticles \cite{BVprl}. 

The order parameter of $^3$He-B in the magnetic field is anisotropic.  The
orientation of the orbital anisotropy axis $\hat{\bf l}$ slowly varies in
the sample (forms a texture) which creates a trapping potential for magnon
quasiparticles in the radial direction owing to the spin-orbit interaction
energy:
\begin{equation}
F_{\rm so} = \frac45 \hbar \frac{\Omega_B^2}{\omega_{\rm L}} 
\sin^2\frac {\beta_l}{2} \, |\Psi|^2\,.
\label{eq:so}
\end{equation}
Here $\omega_{\rm L} = \gamma H$ is the Larmor frequency, $\Omega_B$ is the
Leggett frequency in the B phase, $\beta_l$ is the deflection angle of $\hat{\bf l}$ from the vertical direction (growing
from $\beta_l = 0$ at the cylinder
axis to $\beta_l = \pi/2$ at the cylindrical wall) and $\Psi$
is the wave function of the magnon condensate. Density of magnons
$|\Psi|^2$ is related to the tipping angle of magnetization $\beta_M$ as
$|\Psi|^2 = \chi H (1-\cos\beta_M)/\gamma\hbar$.

Trapping in the axial direction
is provided by an additional pinch coil which creates a minimum in the static
NMR field $H$ and thus minimum in the Zeeman energy
$F_{\rm Z} = \hbar \omega_{\rm L} |\Psi|^2$. The lowest magnon levels in
this trapping potential typically closely follow harmonic-trap relation
\cite{magprl} and 
can be
enumerated by the radial and axial quantum numbers $m$ and $n$,
respectively:
\begin{equation}
f_{mn} = f_{\rm L} + \nu_r(m+1) + \nu_z(n+1/2)\,.
\label{eq:harm}
\end{equation}
Here $f_{\rm L} = \omega_{\rm L}/2\pi \approx 0.826\,$MHz and $\nu_r\approx
220\,$Hz and $\nu_z\approx40\,$Hz are the radial and axial trapping
frequencies, respectively. When magnons are pumped to the trap using NMR
they relax in sub-second time to the ground level \cite{magprl}, where
spontaneous coherence appears and a Bose-Einstein condensate is formed. The
magnetization of the condensate precesses around the magnetic field
at $f_{00} = f_{\rm L} + \nu_r + \nu_z/2$ frequency. The precession induces
signal in the NMR pick-up coil from which the frequency of the precession can
be determined. The free surface changes $\nu_z$ by limiting the trap
in the axial direction and also modifies $\nu_r$ owing to orientation of
$\hat{\bf l}$ perpendicular to the surface. When the waves modify the
geometry of the surface, the frequency of the precession $f_{00}$ changes
as a result.

In the measurements we
keep a small cw pumping usually at $m=2$ level to compensate for the loss
of the magnons from the ground level.  Such pumping is applied at the
frequency $f_{20} > f_{00}$ and thus it does not interfere with the
measurements of the precession of the ground-level condensate.
To the signal recorded from the NMR pick-up coil we apply the band-pass filter to keep only the
contribution from the ground-state condensate including all the side bands,
resulting from the frequency modulation. The frequency $f_{00}$ of the
precession of the condensate is found from the
time intervals between zero crossings in the filtered signal.

\begin{figure}
\centerline{\includegraphics[width=0.9\linewidth]{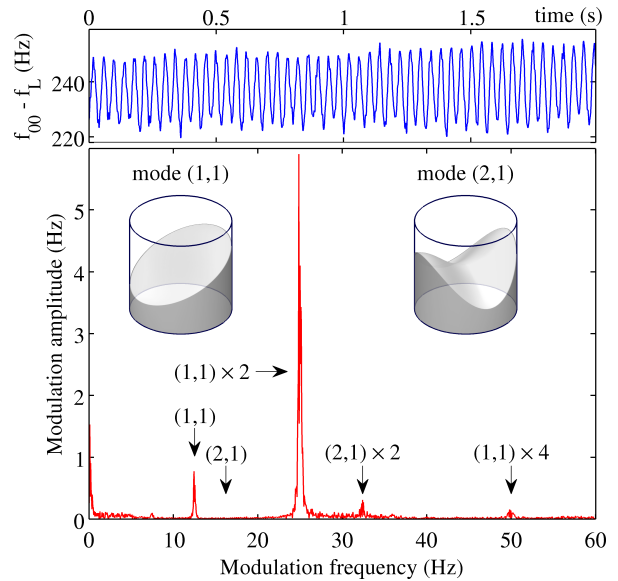}}
\caption{Frequency modulation of the NMR precession owing to
  the  surface waves. \textit{(Top)} Frequency of the precession of the magnon
  BEC in a trap bordering the free surface of the $^3$He-B sample as a function of
  time. The signal is recorded at $T = 0.15\, T_{\rm c}$ with only residual
  vibrations of the cryostat. Modulation of the precession
  frequency is caused by periodic distortion of the trapping potential
  by the surface waves. \textit{(Bottom)} Spectrum of the
  frequency modulation of the NMR precession.
  The plot shows Fourier transform of the signal like in the upper panel,
  but measured for 20\,s. Peaks corresponding to the two surface wave
  modes, cartooned in the inserts, can be clearly identified and are marked
  with arrows. Harmonics are shown by the $\times n$ signs.}
\label{freqspecfig}
\end{figure}

\textit{Surface resonances.}
An example of the measured $f_{00}(t)$ record and its Fourier transform
are shown in Fig.~\ref{freqspecfig}. Peaks in
the frequency
modulation spectrum can be attributed to
the lowest-frequency gravity-wave modes in a vertical cylinder. The height
profile $h(r,\theta)$ of such surface oscillations is
\begin{equation}
h(r,\theta) = J_i(k_{ij} r) e^{i \theta},\; i = 0,1,\ldots, j = 1,2,\ldots
\label{eq:h}
\end{equation}
Here $(r,\theta)$ are the polar coordinates of the point on the surface,
$J_i$ are the Bessel functions, and wave numbers $k_{ij}$ satisfy the
equation
$J'_i(k_{ij} R) = 0$.
The spectrum of these modes follows simple relation for the gravity waves
on deep water
$\omega_{ij}^2 = g k_{ij}$,
where $g$ is the free-fall acceleration.  This applies since the length of
the sample cylinder significantly exceeds its diameter and the surface
tension of $^3$He is small and can be neglected here \cite{surftens}. 

The primary mode is
the non-axisymmetric (1,1) mode with $k_{11} = 1.8412/R$ and its frequency
in our cylinder is $\omega_{11}/2\pi = 12.4\,$Hz. The next lowest-frequency
mode has $k_{21} = 3.0542/R$ and $\omega_{21}/2\pi = 17.8\,$Hz. These two
modes are clearly seen in Fig.~\ref{freqspecfig} with the second harmonic of
the (1,1) mode being the most prominent peak. The frequency doubling occurs
since the frequency of the magnon precession
is the same for the two trap
configurations symmetric relative to the vertical plane passing through the
nodal line of the surface mode. 

The amplitude of the surface waves $h_0$ can be connected to the change of
the frequency of the magnon precession $\Delta f_{00}$ with
Eq.~(\ref{eq:so}) as $\hbar \cdot 2\pi\Delta
f_{00} \sim \hbar (\Omega_B^2/\omega_{\rm L}) (h_0/R)^2$. With $\Delta
f_{00} = 50\,$Hz and $\Omega_B/2\pi \approx 10^5\,$Hz 
we get for the amplitude of the wave $h_0 \sim 0.2\,$mm. At
such amplitudes the flow velocity along the surface $v \sim (kR) h_0
\omega/\pi \lesssim 1\,$cm/s is substantially below the critical and we have not observed formation of vortices by surface oscillations.

\begin{figure}[t]
\centerline{\includegraphics[width=0.9\linewidth]{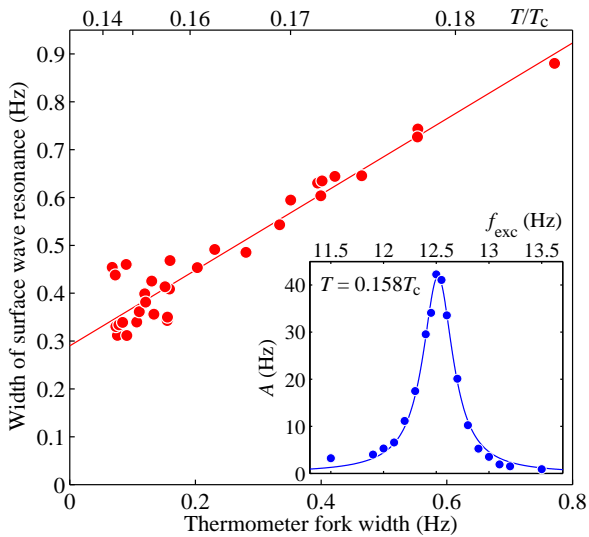}}
\caption{Damping of the surface waves as a function of temperature. The
  measured width of the primary surface wave resonance $\Delta f_{\rm
    surf}$ is plotted with circles against vertical axis.  Bottom
  horizontal axis shows the resonance width of the quartz tuning fork used
  as a thermometer \cite{forktherm}. Corresponding temperature values are shown on the upper
  axis. Solid line is a linear fit to data points. \textit{(Insert)} An
  example of a measurement of the surface wave resonance. Surface
  oscillations are mechanically excited at frequency $f_{\rm exc}$ and the
  amplitude of the frequency modulation $A$ in the precession of the magnon
  condensate is measured (circles). Solid line shows a fit to the
  square of the standard Lorentzian response, from which $\Delta f_{\rm
    surf}$ is determined.}
\label{tempdepfig}
\end{figure}

\textit{Surface damping.} To determine the damping of the surface waves we
have measured the resonance width $\Delta f_{\rm surf}$ of the forced
oscillations in the primary mode, Fig.~\ref{tempdepfig}. Oscillations are
excited by periodically tilting the cryostat at a frequency $f_{\rm exc}$
using the active air-spring dampers on which the cryostat is suspended.
With the vertical distance from the suspension point to the sample surface
of about 1.5\,m this transforms essentially to horizontal oscillations of
the sample container. We measure $f_{00}(t)$ dependence like in
Fig.~\ref{freqspecfig}(top) and extract the amplitude of the response $A$
at $2f_{\rm exc}$ frequency using a lock-in-like detection.

As seen in Fig.~\ref{tempdepfig} the damping increases with increasing temperature
as the density of the normal component grows. We are not aware of the rigorous calculation
of the damping applicable at these temperatures, where the
normal component of $^3$He-B presents a gas of ballistic quasiparticles, which would include also the
effects of Andreev reflection from the surface \cite{surfandreev} and from
the flow along it. A simple model of thermal damping of quartz tuning forks
in the ballistic regime \cite{lancvibr} predicts that the width of the
resonance is proportional to $(d/M) \exp(-\Delta/T)$, where $d$ is the size
of the object perpendicular to the direction of the oscillations, $M$ is
the effective mass and the exponential factor reflects
temperature-dependent density of thermal quasiparticles. Using this
expression we can roughly scale the thermal effect from the thermometer
fork, which has $d/M
\approx 270\,$cm/g \cite{ourfork}, to the oscillations of the surface. For the primary mode of the oscillating surface we
estimate $d\sim R$ and $M \sim \rho_{\rm He} R^2/k$ which gives $d/M \sim
200\,$cm/g. Thus we may expect that the thermal contribution to the width
of the resonant response will be of the same order for our quartz tuning
forks and for the oscillating surface. This is indeed demonstrated by the
data in Fig.~\ref{tempdepfig}, where the slope of the fit line is close to
1. We note, however, that this slope has not been exactly reproducible in the runs
which differ by the azimuthal orientation of the cryostat (and thus
different directions of the forcing with respect to the residual
misalignment of the sample axis and the vertical direction), which is
probably related to the sensitivity of the surface wave pattern in a
vibrating cylinder to the exact conditions of the forcing \cite{miles2}.

Another feature seen by Fig.~\ref{tempdepfig} is the
finite value of the surface resonance width (0.3\,Hz) when extrapolated to zero
temperature. This zero-temperature damping has been
reproducible in all measurement runs. Among possible explanations for this
damping is the surface friction at the cylindrical wall of the sample or
non-linear interactions with other surface wave modes and possible
creation of wave turbulence \cite{gravwaveturb,waveturb4he}. An intriguing possibility is 
contribution to the damping from the surface-bound Majorana fermions. No calculations of such contribution
exist up to date, though, and even the physical mechanism of possible
damping is not entirely clear. It can be similar to the
recently proposed temperature-independent but frequency-dependent
dissipation mechanism in the motion of quantized vortices originating from
the vortex-core-bound fermions \cite{silaev}. Alternatively surface-bound fermions can directly
mediate energy and momentum transfer to the container walls without
transferring them to the bulk quasiparticles first. Such contribution to
damping should have power-law dependence on temperature which in our
experimental temperature range would mimic a finite zero-temperature
damping.

\begin{figure}[t]
\centerline{\includegraphics[width=0.9\linewidth]{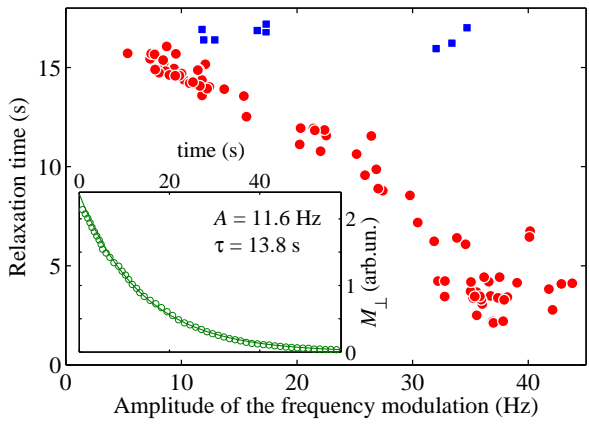}}
\caption{Relaxation of the magnon condensate in a
  periodically modulated trap. Relaxation time is plotted as a function of
  the amplitude $A$ of the modulation of the ground level in the
  trap. Circles show the modulation owing to the surface waves and
  squares modulation with the trapping magnetic field. Modulation frequency
  in both cases is 25\,Hz. \textit{Insert}
  presents an example of a measurement of the relaxation time $\tau$. The
  amplitude of the magnon condensate signal in the NMR pick-up coil, which
  is proportional to the transverse magnetization $M_\perp$, is shown as a
  function of time after switching the pumping off (circles). The line is an
exponential fit $M_\perp = M_{\perp0} \exp(-t/\tau)$ from which $\tau$ is
determined. }
\label{relaxfig}
\end{figure}

\textit{Relaxation of the magnon BEC.}
Additional information about the surface waves is provided
by the measurements of the relaxation of the magnon condensates after
switching the NMR pumping off.
For the magnon condensate in a time-independent trap which is completely in bulk the relaxation is
determined by the bulk thermal quasiparticles \cite{lancrelax} and by the interaction
with a NMR pick-up circuit, which will be described elsewhere. For our
condensates in a time-dependent potential modulated by the surface waves we
observe an additional relaxation channel so that the life time of magnon
condensates decreases with increasing
amplitude of the surface oscillations, Fig.~\ref{relaxfig}. 

In principle, for a quantum-mechanical system in a time-dependent potential
one can expect transitions to other states. However, for our trapped
condensates in the ground state we expect the highest probability of
excitation at modulation frequencies corresponding to transitions between
different levels in the trap. Such resonant depletion was observed in
Bose-Einstein condensates of cold atoms \cite{becreson}. In our case the
modulation of the trapping potential is at the frequency
$\omega_{11}/\pi\approx 25\,$Hz while the lowest excited level is at
$2\nu_z \approx 80\,$Hz and thus the resonant conditions are not satisfied.
We have also measured the relaxation rate in the case when the trapping
potential is periodically modulated with the same 25\,Hz frequency using
the current in the minimum field coil. In such a case no enhancement of the
relaxation has been observed (squares in Fig.~\ref{relaxfig}). 

Thus it seems likely that the
modulation of the trapping potential in itself can not explain the effect
of the surface waves on the magnon relaxation and one may wonder whether this effect
is related to the surface-bound Majorana fermions. It has been suggested
\cite{elrelax} that Majorana fermions can be probed with the relaxation of
the spin of an electron, which is localized in a bubble close to the surface
of $^3$He-B. Magnon condensates have significantly larger
coherently precessing spin than a single electron and so should be more
sensitive to the surface relaxation effects. Oscillating surface might
enhance relaxation processes in the bound-states subsystem which would
allow for a faster energy transfer from the condensate to the bound states.

\textit{To conclude,} we have observed gravity waves on a free surface of the
ultra-cold superfluid $^3$He-B. In the temperature range below $0.2T_{\rm
  c}$ the waves are only weakly damped and are easily excited by minute
vibrations of the sample container. With decreasing temperature the damping
decreases linearly with the density of bulk thermal quasiparticles, but
extrapolates to a finite value in the $T \rightarrow 0$ limit. The
relaxation rate of magnon Bose-Einstein condensates in a magneto-textural
trap attached to the surface is enhanced by the surface waves, but is
not affected by similar
modulation of the trapping potential with the magnetic
field. To establish
a possible link between these observations and Majorana fermions bound to
the surface of $^3$He-B a theoretical model of the relevant phenomena
should be built. An interesting development will be to study the
shallow-water case, where surface waves with relativistic spectrum can be
used to construct analogues of black holes \cite{bholes} including
Hawking radiation \cite{hawkinganal}. Ultra-cold $^3$He-B
has no viscosity and provides coupling of the waves to the fermionic quantum
vacuum of the superfluid, which should make such analogue systems much
richer compared to water waves. Another possibility is to use similarities
between the magnon BEC confined with the $\hat{\bf l}$ texture field and
the confined quarks in the MIT bag model \cite{mitbaganalog} to study
'high-energy physics' on a moving brane, role of which is played by the surface
of $^3$He.

We thank G.E.~Volovik, M.~Krusius, M.A.~Silaev
and I.A.~Todoshchenko for stimulating discussions. The work is supported by
the Academy of Finland (Center of Excellence 2012-2017) and the
EU 7th Framework Programme (grant 228464 Microkelvin).
P.J.H. acknowledges financial support from the V\"{a}is\"{a}l\"{a}
Foundation.


\begin{thebibliography}{99}

\bibitem{LL} L.\,D. Landau and E.\,M. Lifshitz, {\it Fluid Mechanics}, 2nd
  ed., Elsevier, Amsterdam, 1987, \S\S 12 and 25.

\bibitem{surfnstar} A.\,L. Piro and L. Bildsten,
Astrophys. J. \textbf{629}, 438 (2005).

\bibitem{surf4he} T.  Takahashi et al., Microgravity Sci. Technol.
 \textbf{23}, 365 (2011).

\bibitem{oilfilm} J.\,C Scott, Nature
  \textbf{340}, 601 (1989).

\bibitem{he3majorana} G.\,E. Volovik, Pis'ma
  v ZhETF \textbf{90}, 440 (2009) [JETP Lett. \textbf{90}, 398 (2009)].

\bibitem{topsuper} X.-L. Qi, T.\,L. Hughes, S. Raghu, and S.-C. Zhang,
Phys. Rev. Lett. \textbf{102}, 187001 (2009).

\bibitem{elrelax} S.\,B. Chung and S.-C. Zhang, Phys. Rev.  Lett.
  \textbf{103}, 235301 (2009).

\bibitem{okudarev} Y. Okuda, and R. Nomura,
J. Phys.: Condens. Matter \textbf{224}, 343201 (2012).

\bibitem{majoranasss} V. Mourik et al., Science \textbf{336}, 1003 (2012).

\bibitem{3rdsound} A.\,M.\,R. Schechter, R.\,W. Simmonds, R.\,E. Packard
  and J.\,C. Davis, Nature {\bf 396}, 554
  (1998).

\bibitem{BVprl} Yu.\,M. Bunkov and G.\,E. Volovik, Phys. Rev. Lett.
  \textbf{98}, 265302 (2007).

\bibitem{magprl} S. Autti et al., Phys. Rev. Lett.  \textbf{108}, 145303
  (2012).

\bibitem{surftens} K. Matsumoto, Y. Okuda, M. Suzuki and S. Misawa, J. Low Temp. Phys. {\bf 125}, 59
  (2001).

\bibitem{surfandreev} T. Okuda, H. Ikegami, H. Akimoto and H. Ishimoto,
  Phys. Rev. Lett.  \textbf{80}, 2857 (1998).

\bibitem{lancvibr} D.\,I. Bradley et al., J.
  Low Temp. Phys. \textbf{157}, 476 (2009).

\bibitem{ourfork} The thermometer fork has tines with dimensions
0.1\,mm$\times$0.24\,mm$\times$2.4\,mm.


\bibitem{miles2} J.\,W.\,Miles, J. Fluid Mech. {\bf 149}, 15 (1984).

\bibitem{gravwaveturb} P. Denissenko, S. Lukaschuk, and S. Nazarenko, Phys.
  Rev. Lett. \textbf{99}, 014501 (2007).

\bibitem{waveturb4he} L.V. Abdurakhimov, M.Yu. Brazhnikov, A.A. Levchenko,
  I.A. Remizov, and S.V. Filatov,
Physics-Uspekhi \textbf{55}, 818 (2012).


\bibitem{silaev} M.\,A. Silaev, Phys. Rev. Lett.  \textbf{108}, 045303 (2012).

\bibitem{lancrelax} S.\,N. Fisher, G.\,R. Pickett, P. Skyba, and N. Suramlishvili , Phys. Rev. B {\bf 86}, 024506 (2012).

\bibitem{becreson} D. Hunger et al., Phys.
  Rev. Lett.  \textbf{104}, 143002 (2010).

\bibitem{bholes} R. Sch\"utzhold and W.\,G. Unruh, Phys. Rev. D
  \textbf{66}, 044019 (2002).

\bibitem{hawkinganal} S. Weinfurtner, E.\,W. Tedford, M.\,C.\,J. Penrice,
  W.\,G. Unruh, and G.\,A. Lawrence,
 Phys. Rev. Lett. {\bf 106}, 021302 (2011).

\bibitem{mitbaganalog} S. Autti, V.\,B. Eltsov, and G.\,E. Volovik,
Pis'ma v ZhETF \textbf{95} 611 (2012) [JETP Lett. \textbf{95}, 544 (2012)].

\bibitem{forktherm} R. Blaauwgeers et al., J. Low Temp. Phys. {\bf 146},
  537 (2007).



\end{thebibliography}
\end{document}